\documentclass[aps,prb,preprint,amsmath,amssymb,superscriptaddress,floatfix,notitlepage]{revtex4-2}
\usepackage{graphicx}
\usepackage[colorlinks, citecolor=blue, linkcolor=blue]{hyperref}
\usepackage{xcolor}
\usepackage{float}
\usepackage{setspace}
\usepackage[font=small,format=plain,labelfont=bf,
singlelinecheck=false,justification=raggedright,
labelsep=space]{caption}
\usepackage[normalem]{ulem}
\definecolor{darkorchid}{HTML}{0000CD}
\definecolor{darkpink}{HTML}{880e4f}

\newcommand{\tq}{turquoise}

\newcommand{\KIEL}{Institute of Theoretical Physics and Astrophysics, University of Kiel, Leibnizstrasse 15, 24098 Kiel, Germany}
\newcommand{\KINSIS}{Kiel Nano, Surface, and Interface Science (KiNSIS), University of Kiel, 24118 Kiel, Germany}
\newcommand{\HAM}{Institute of Nanostructure and Solid State Physics (INF), University of Hamburg, Jungiusstraße 11, 20355 Hamburg, Germany}

\begin{document}
\title{Kicking Co and Rh atoms on a row-wise antiferromagnet}

\author{Felix Zahner}\affiliation{\HAM}
\author{Soumyajyoti Haldar}\email[]{haldar@physik.uni-kiel.de}\affiliation{\KIEL}
\author{Roland Wiesendanger}\affiliation{\HAM}
\author{Stefan Heinze}\affiliation{\KIEL}\affiliation{\KINSIS}
\author{Kirsten von Bergmann}\affiliation{\HAM}
\author{André Kubetzka}\email[]{kubetzka@physnet.uni-hamburg.de}
\affiliation{\HAM}

\date{\today}

\begin{abstract}
Diffusion on  surfaces is a fundamental process in surface science, governing nanostructure and film growth, molecular self-assembly, and chemical reactions. Atom motion on non-magnetic surfaces has been studied extensively both theoretically and by real-space imaging techniques. For magnetic surfaces density functional theory (DFT) calculations have predicted strong effects of the magnetic state onto adatom diffusion, but to date no corresponding experimental data exists. Here, we investigate Co and Rh atoms on a hexagonal magnetic layer, using scanning tunneling microscopy (STM) and DFT calculations. Experimentally, we ``kick'' atoms by local voltage pulses and thereby initiate strictly one-dimensional motion which is dictated by the row-wise antiferromagnetic (AFM) state. Our calculations show that the one-dimensional motion of Co and Rh atoms results from conserving the Co spin direction during movement and avoiding high induced Rh spin moments, respectively. These findings demonstrate that magnetism can be a means to control adatom mobility.
\end{abstract}

\maketitle

\begin{figure}
    \centering
    \vspace{5mm}
    \includegraphics[width=0.5\linewidth]{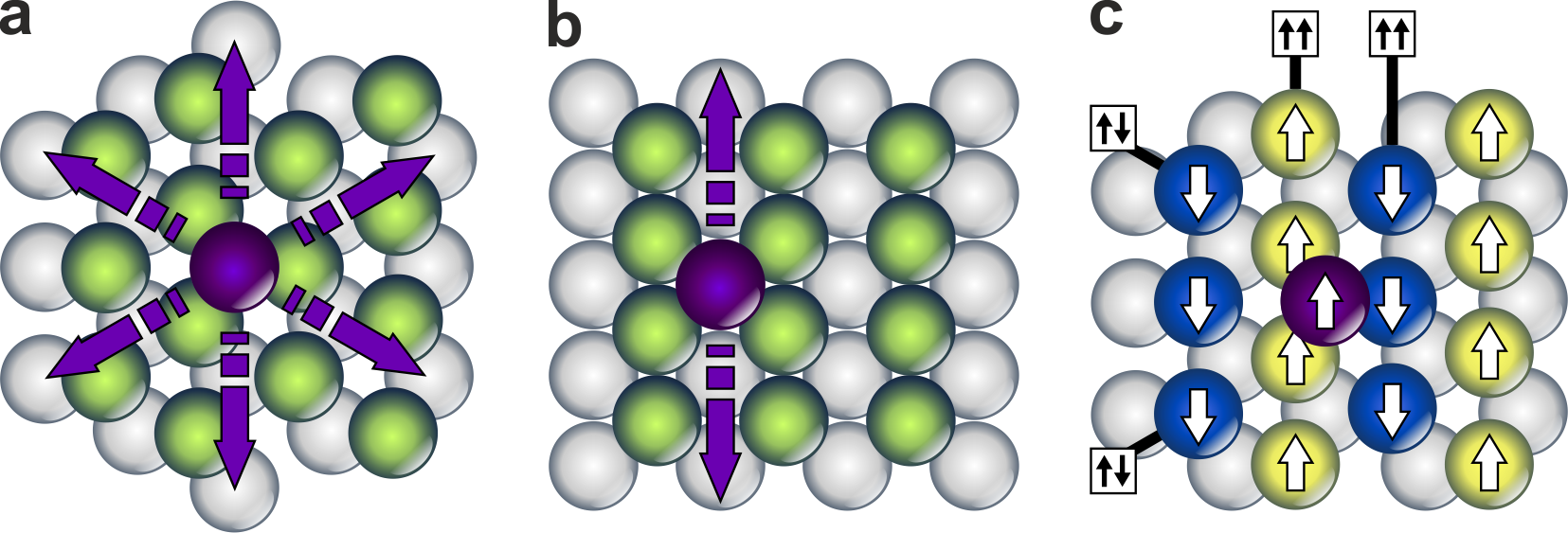}
    \caption{\textbf{| Structural and magnetic surface symmetry. a,} Quasi-isotropic diffusion of an adatom on a hexagonal surface with six structurally equivalent directions. \textbf{b,} Fcc(110) surface with typically preferred diffusion along close-packed atomic rows \cite{Li1996,Li1998}.  \textbf{c,} Adatom on a row-wise antiferromagnet, with the  $\uparrow\uparrow$-rows breaking the hexagonal surface symmetry. Its magnetic moment is assumed to be parallel to the majority of neighboring Mn surface atoms.}
    \label{fig:start}
\end{figure}

The movement of atoms and molecules on surfaces is a fundamental and well-studied process in surface science. Adatoms typically move thermally driven by random jumps between neighboring binding sites, while longer jumps and exchange processes with the surface are activated with increasing temperature \cite{Antczak2007}.
The surface symmetry plays a key role for these processes, see Fig.\,\ref{fig:start}a,b: whereas diffusion is effectively isotropic on hexagonal surfaces, lower symmetry surfaces like fcc(110) can show strong directional anisotropies \cite{Frenken1990,Li1996}, or even  strictly one-dimensional movement \cite{Swartzentruber1995}. Magnetic effects are usually neglected, either because diffusion is studied above the magnetic ordering temperature, $T_{\rm C}$, or because their strengths are considered negligible compared to binding energies and diffusion barriers.

For bulk systems, however, experiments indicate that magnetic states can affect diffusion. In bcc iron, for example, the self-diffusion coefficient systematically deviates from a linear Arrhenius relation, i.e.\  below $T_{\rm C}$ Fe atoms diffuse more slowly through the Fe crystal than extrapolated from higher temperatures \cite{Iijima1988}. In some alloy systems, the structural composition can be influenced by applying external magnetic fields during annealing; for the antiferromagnet PdMn it has been argued that excess Pd atoms preferentially diffuse to the magnetic sublattice which is favored by the external field
\cite{Josten2023}. 

On magnetic surfaces, so far no diffusion experiments have been performed, but STM manipulation experiments have reported alternating forces, when a magnetic atom is moved across an antiferromagnetic surface with a magnetic tip \cite{Wolter2012,Spethmann2021}. On the theory side, almost 20 years ago DFT calculations for self-diffusion on Fe and Mn monolayers on W(110) and W(001) have predicted a strong influence of magnetism onto adsorption sites and diffusion properties with a general trend of enhanced diffusion on ferromagnetic surfaces \cite{Spisak2004,Dennler2005}, but respective experiments are still missing.

Here, we investigate how Co and Rh atoms move at $T=4.2$~K on a hexagonal Mn layer hosting a row-wise AFM state, when they are ``kicked'', i.e.\ when motion is initiated by local voltage pulses from an STM tip. For both adatom species we observe strictly one-dimensional movement along the rows of Mn atoms with parallel spins ($\uparrow\uparrow$-rows), see Fig.\,\ref{fig:start}c. DFT calculations show that the pseudomorphic Mn layer is shifted laterally by 15~pm off the fcc hollow site positions on Re(0001), i.e.\ the Mn layer and adatom $C_3$ (threefold) symmetry is broken by the magnetic state electronically as well as structurally. For both adatom species the energy barriers differ by about 50~meV for diffusion across Mn bridge sites with the two Mn atoms having the same ($\uparrow\uparrow$-bridge) compared to opposite spin orientation ($\uparrow\downarrow$-bridge), see Fig.\,\ref{fig:electronic}c, making the potential landscape highly anisotropic.


\section{Experimental Results}
\begin{figure*}
    \centering
        \includegraphics[width=0.9\textwidth]{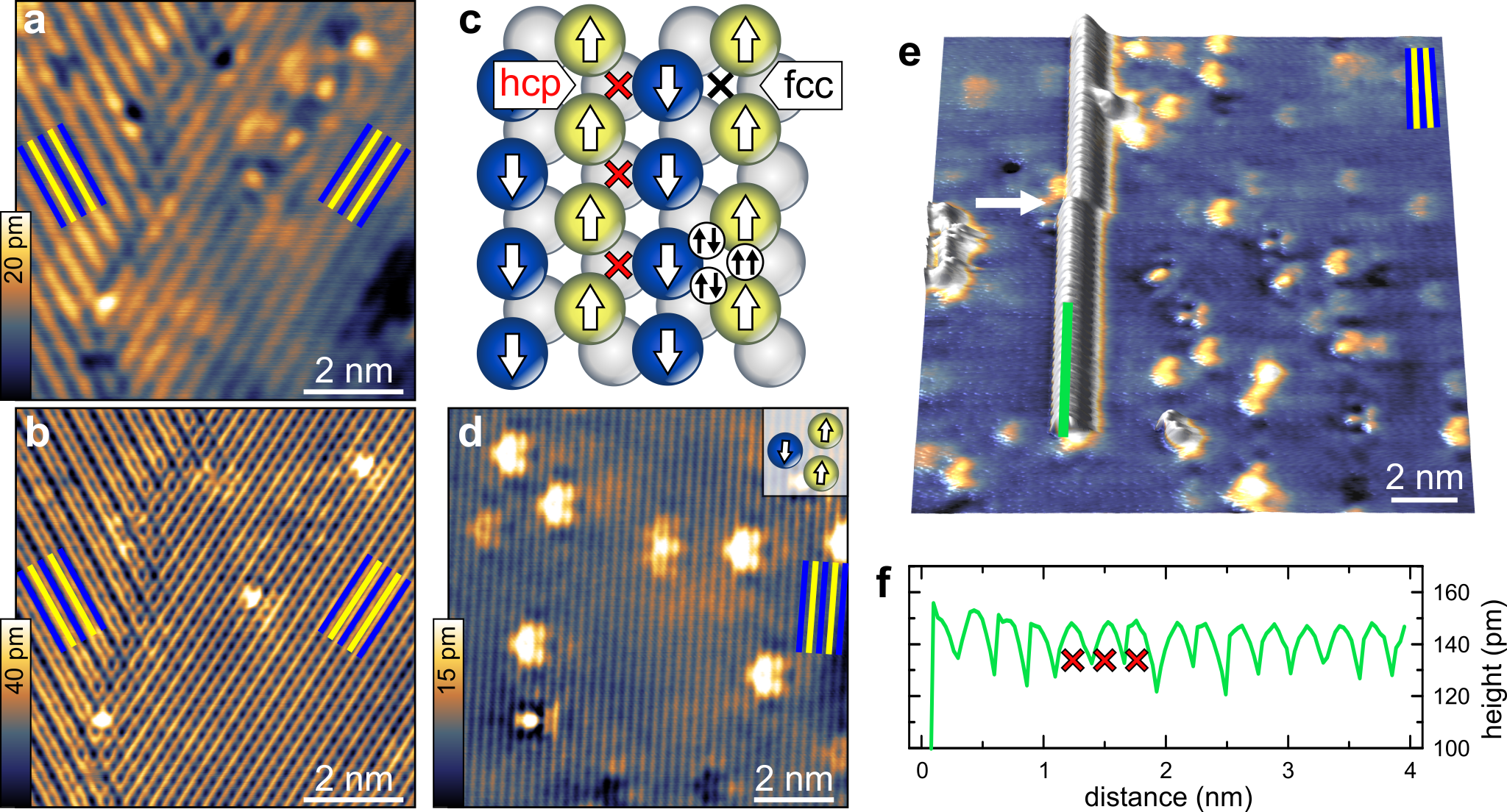}
    \caption{\textbf{| Broken surface symmetry and one-dimensional Co atom movement. a,} Two rotational domains of the row-wise AFM state in fcc-Mn/Re(0001), imaged by constant current SP-STM ($U=+14$~mV, $I=1$~nA). \textbf{b,} With the tip closer to the surface ($U = +5$~mV, $I = 70$~nA),  additional stripes of atomic period appear. \textbf{c,} Sketch of the magnetic state indicating different hollow sites (red and black crosses) and inequivalent $\uparrow\uparrow$- and $\uparrow\downarrow$-bridge sites.
    \textbf{d,} High resolution non-magnetic STM image ($U=+30$~mV, $I=1$~nA) of the $\uparrow\uparrow$-rows with native defects reflecting the broken symmetry of the hollow site, see inset. \textbf{e,}  A single Co atom is imaged while it is moving to the top of the image, guided by the $\uparrow\uparrow$-rows (slow scan direction: bottom$\rightarrow$top, fast scan direction: left$\rightarrow$right).    \textbf{f,} The line section shows that the Co atom is imaged with atomic periodicity of 2.7~\AA, see red crosses in \textbf{c}.}
    \label{fig:electronic}
\end{figure*}

 The row-wise AFM state within the hexagonal Mn layer on Re(0001) \cite{Spethmann2020} allows for three equivalent rotational magnetic domains, two of which are seen in the spin-polarized (SP) STM data of Fig.\,\ref{fig:electronic}a. Due to spin-polarized tunneling, neighboring atomic rows display different signal strengths \cite{Wortmann2001} and the left domain exhibits a stronger magnetic contrast for this particular tip magnetization direction \cite{Spethmann2021}. At lower tunnel voltage and smaller tip--sample distance (higher current) in Fig.\,\ref{fig:electronic}b
 the hexagonal atomic lattice becomes visible and an additional stripe pattern appears with half the period of the magnetic state. Since all Mn atoms are electronically equivalent, this atomic stripe pattern must stem from the inequivalent $\uparrow\uparrow$- and $\uparrow\downarrow$-bridge sites, see Fig.\,\ref{fig:electronic}c, indicating that the $C_3$ symmetry of the Mn hollow sites is broken.
 For the same reason many native defects display a boomerang-like shape which is correlated with the magnetic state, see Fig.\,\ref{fig:electronic}c,d and Methods. This defect shape, which has only one mirror plane, reflects a loss of electronic $C_3$ surface symmetry and allows to determine the orientation of the magnetic domains with a non-magnetic STM tip. Fig.\,\ref{fig:electronic}d shows different kinds of these defects with high resolution as well as the $\uparrow\uparrow$-rows of the magnetic state. In this non-magnetic data, neighboring $\uparrow\uparrow$-rows appear identical.

  With a different non-magnetic tip in Fig.\,\ref{fig:electronic}e, the Mn layer itself appears featureless, but the direction of the $\uparrow\uparrow$-rows can still be inferred from the defect shapes. Co adatoms on fcc-Mn/Re(0001) are easily moved during standard STM imaging: in Fig.\,\ref{fig:electronic}e a single Co atom jumps along the $\uparrow\uparrow$-rows, following the slow scan direction of the tip upward, and it therefore appears like an atomic chain. Only once does the atom jump one atomic row to the right (see arrow in Fig.\,\ref{fig:electronic}e), flipping its magnetic moment in the process, see Fig.\,\ref{fig:start}c. This data is reminiscent of the experiments performed by Li {\em et al.}\ with Ag atoms on Ag(110) \cite{Li1996,Li1998} and is a first hint that the magnetic state dictates the movement direction of Co atoms. The line section in Fig.\,\ref{fig:electronic}f shows the period of the atomic lattice, indicating that only one type of hollow site is a stable position for Co, in agreement with no zigzag movement being observed. On all other positions, the residence time is too short to be detectable in this measurement. We will see in section \ref{section:theory} that the Co atoms prefer the so-called hcp hollow site above the surface Re atom, see red crosses in Fig.\,\ref{fig:electronic}c.

\begin{figure*}
    \centering
    \includegraphics[width=0.5\linewidth]{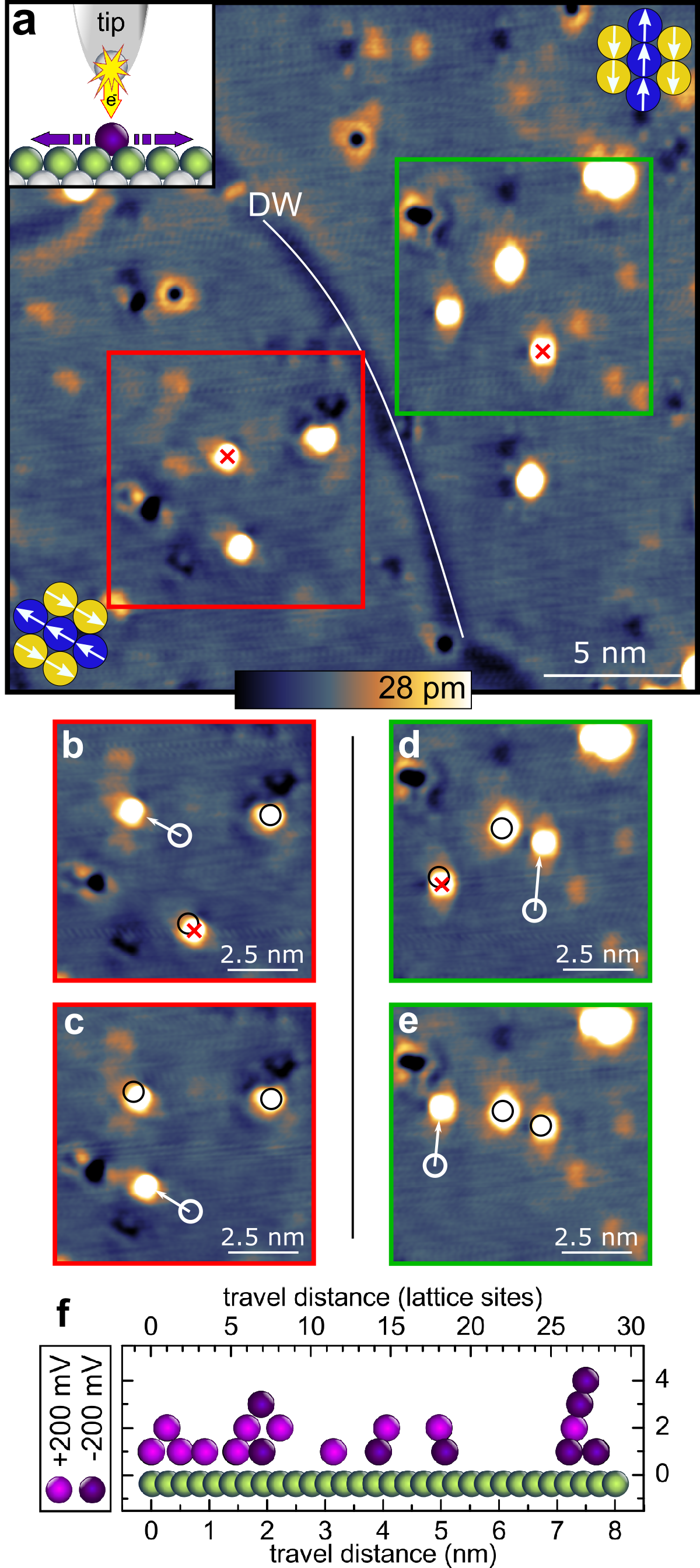}
    \caption{{\setstretch{1.2}\textbf{| Kicking Co atoms on a row-wise antiferromagnet. a,} Overview STM image of Co atoms on fcc-Mn/Re(0001) with two rotational domains; their orientation is apparent by the defect shapes and the domain wall (DW) is imaged via a reduced electron density. Kicking is performed by a sudden voltage raise at constant tip height, see inset. \textbf{b,} The upper left Co atom has moved along the $\uparrow\uparrow$-rows after kicking. \textbf{c,} The bottom Co atom has moved in the same direction after kicking. \textbf{d,} The bottom Co atom has moved along the $\uparrow\uparrow$-rows after kicking. \textbf{e,} The left Co atom has moved in the same direction after kicking. (All images $U=+8$~mV, $I=1$~nA, all voltage pulses: $U_\text{P}=+200$~mV, $\Delta t=1$~s, tip kicking locations are marked by crosses in the previous image, previous atom positions are marked by circles.) \textbf{f,} Kicking distances for both voltage polarities (color-coded), $|U_\text{P}|=200$~mV, $\Delta t=0.5$~s.}}
    \label{fig:Co}
\end{figure*}

To minimize the influence of the tip, instead of dragging the adatom along during imaging, in the following we ``kick'' the adatoms with a voltage pulse from a stationary tip and determine the new position by subsequent imaging. The kicking is performed by a sudden voltage increase at constant tip height above the atom, which also leads to an increased tunnel current, see inset of Fig.\,\ref{fig:Co}a and Methods. By performing these experiments with one and the same micro-tip on different rotational domains, we can exclude a possible tip asymmetry being responsible for the movement direction.
Figure \ref{fig:Co}a shows a surface area with two rotational domains like in Fig.\,\ref{fig:electronic}a, this time imaged with vanishing spin contrast. The domain wall (DW), now visible via a reduced electron density \cite{Spethmann2021}, is marked by a white line. We find that the Co movement initiated by kicking is always toward the upper left on the left domain and toward the top of the image on the right domain, in both cases following the $\uparrow\uparrow$-rows of the magnetic state (Supplementary Video 1). Examples are given in Fig.\,\ref{fig:Co}b,c (left domain) and Fig.\,\ref{fig:Co}d,e (right domain). Independent of the micro-tip used for kicking, we always observe a strictly one-dimensional movement of Co atoms along the $\uparrow\uparrow$-rows, within an accuracy of one atomic lattice site. Occasionally, Co atoms which are further away from the directly kicked atom  also move 1--2 lattice sites, see Fig.\,\ref{fig:Co}b--d, indicating a weak long range effect. These short jumps also  occur  exclusively along the $\uparrow\uparrow$-rows. 

The Co atoms are surprisingly mobile on fcc-Mn/Re(0001). Depending on the micro-tip, we find a threshold voltage for single-site jumps at rather low voltage, in the range of 8--50~mV at $I=1$~nA. In Fig.\,\ref{fig:Co}b--e Co atoms move 10--15 lattice sites, when kicked by voltage pulses of $U_\text{P}=+200$~mV. A further measurement series of 19 kicking events, using both voltage polarities
(see also section \ref{sec:efield}) is shown in Fig.\,\ref{fig:Co}f: the average travel distance is similar and the distribution is roughly uniform with an accumulation at the far end at 7--8~nm. We observe a maximum kicking distance of about 10~nm at $U_\text{P}=500$~mV. A random walk type motion of uncorrelated short jumps would lead to a normal distribution with a maximum at zero and an exponential tail \cite{Antczak2007}, and can therefore be ruled out. Instead, these results indicate low diffusion barriers along the $\uparrow\uparrow$-rows and an effectively low damping, allowing the Co atoms to perform long jumps \cite{Pollak1993,Ferrando1994}.

\begin{figure*}
    \centering
    \includegraphics[width=0.7\textwidth]{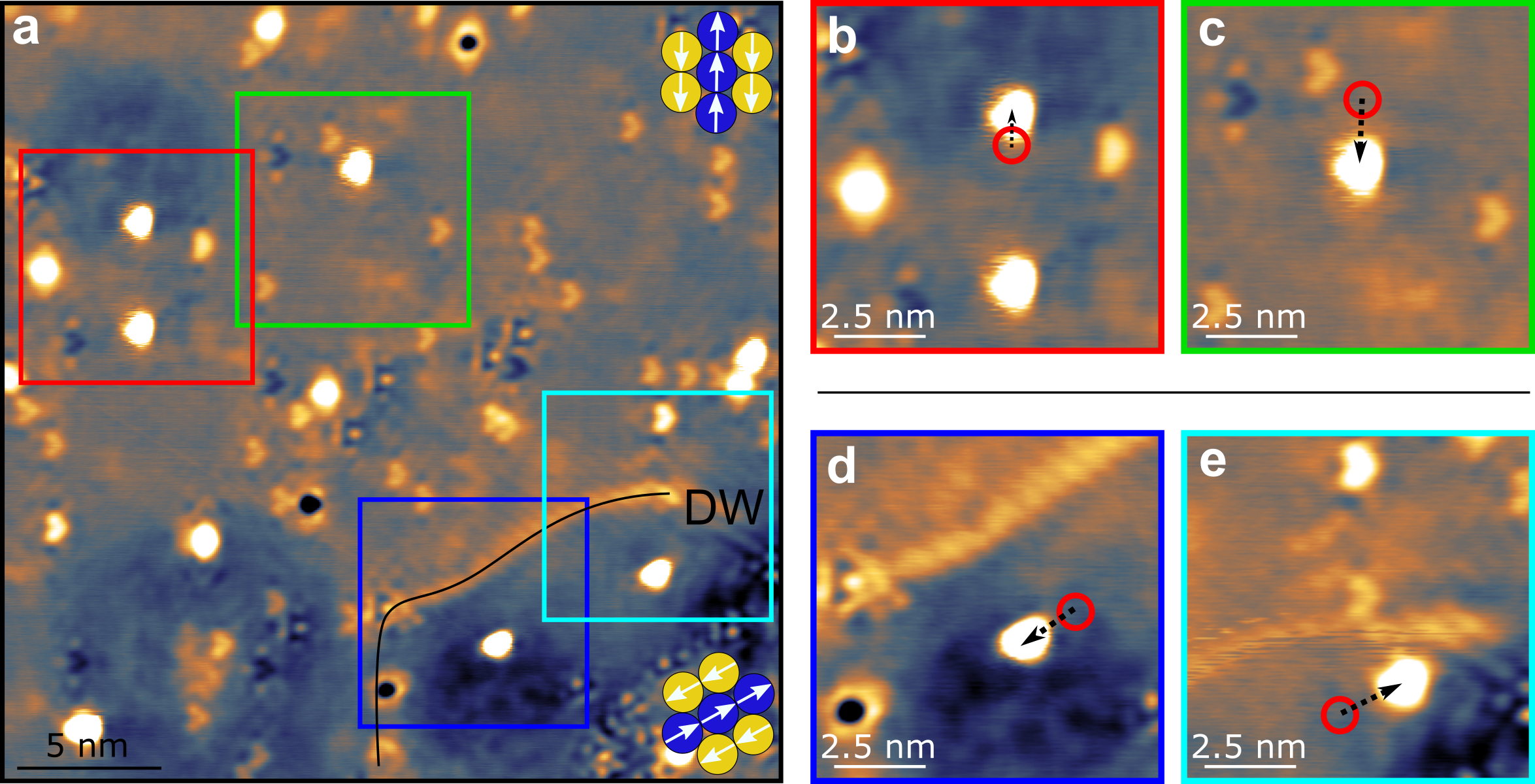}
    \caption{\textbf{| Kicking Rh atoms on a row-wise antiferromagnet. a,} Overview STM image of Rh atoms on fcc-Mn/Re(0001) with two rotational domains. A domain wall (DW) is marked by a black line. \textbf{b--e,} When kicked, the Rh atoms also move according to the rotational domain state, but travel shorter distances compared to Co, despite the use of pulses with higher voltage magnitude. The electron density of native defects and also the Rh atoms is asymmetric due to the AFM state.  (All images $U=+25$~mV, $I=2$~nA, $U_\text{P}=+1$~V, $\Delta t = 400$~ms).}
    \label{fig:Rh}
\end{figure*}

To make a more general case we repeated the above experiments with a second adatom species: Rh is in the same chemical group and thus iso-electronic to Co, but almost 75\% heavier and typically non-magnetic on metal surfaces. We again use a surface area with two rotational domains to exclude tip artifacts, see Fig.\,\ref{fig:Rh}a.
We similarly find that movement is always along the $\uparrow\uparrow$-rows (Supplementary Video 2), as can be seen in Fig.\,\ref{fig:Rh}b,c (upper domain) and in Fig.\,\ref{fig:Rh}d,e (lower domain). The Rh atoms show no spherical symmetry but instead reflect the broken $C_3$ symmetry, like many of the native defects. In comparison, the Co shows an asymmetry in the extended electron density surrounding the atom, see Fig.\,\ref{fig:Co}a--e. Apparently, Rh is much less mobile than Co: at $U_\text{P}=+200$~mV Rh atoms do not move more than one atomic site and at $U_\text{P}=+1$~V they travel only 1.0--2.5~nm, see Fig.\,\ref{fig:Rh}b--e. We observe a maximum travel distance of 3~nm at $U_\text{P}=+2$~V.

For both atom species, magnetic Co and non-magnetic Rh, we thus find strictly one-dimensional movement along the $\uparrow\uparrow$-rows of the AFM state, within an accuracy of one atomic site. The critical parameter to move an atom seems to be voltage magnitude $|U_\text{P}|$. The threshold voltage is tip-dependent and element specific with $U_\text{P}=8$--50~mV for Co and $U_\text{P}=200$--300~mV for Rh.

\section{First-principles calculations}
\label{section:theory}
\begin{figure*}
    \centering
    \vspace{5mm}
    \includegraphics[width=0.9\linewidth]{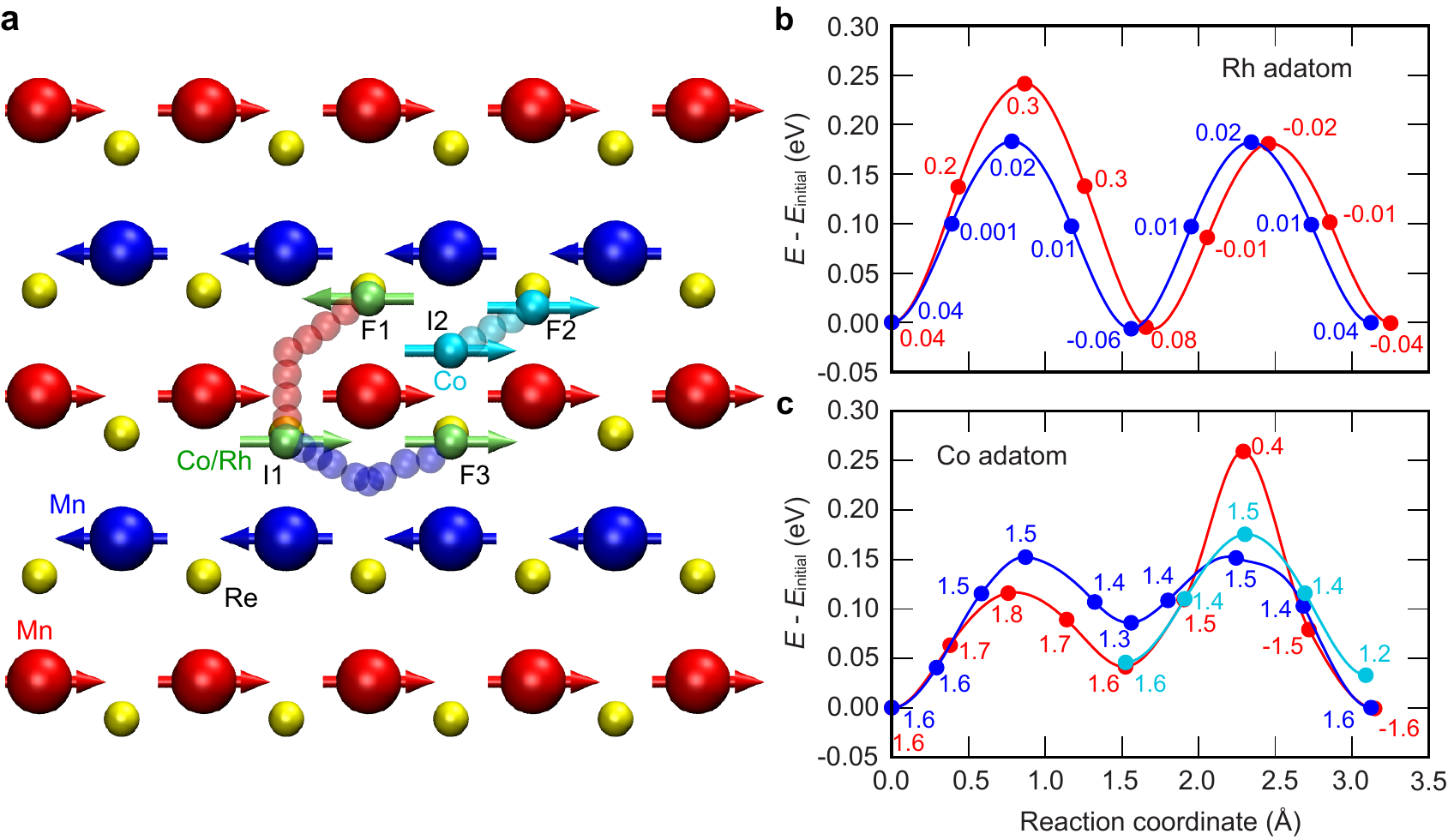}
    \caption{\textbf{| DFT minimum energy paths and energy barriers for Rh and Co atoms on Mn/Re(0001). a,} Minimum energy path for the quasi one-dimensional movement
    of Rh and Co adatoms between close-packed atomic rows of the fcc-Mn/Re(0001) surface in the row-wise AFM state 
    obtained via DFT nudged-elastic band calculations.
    Red and blue spheres with arrows represent Mn atoms with opposite
    spin directions. Yellow spheres denote Re surface atoms, while
    small green and cyan spheres represent adatoms. Note that the lateral shift of the Mn layer
    with respect to the Re surface obtained from DFT is included in the sketch.
    Red and blue transparent spheres indicate
    the calculated paths from the hcp hollow adsorption site (initial
    state 1, I1) to two similar adsorption sites with opposite 
    spin directions (final states F1 and F3).
    For the Co atom the minimum energy path is also given from the
    initial state I2 to the final state F2 (cyan spheres).
    \textbf{b, c} Energy along the minimum energy paths for Co and Rh atoms from
    the initial state I1 to the final states F1 and F3 along the red and blue paths 
    indicated in panel \textbf{a}, respectively. For the Co atom the energy is also
    given along the path from I2 to F2 (cyan symbols and line).
    The magnetic moment of the Co and Rh atom is given in $\mu_{\rm B}$ at every image along the paths.
    }
    \label{fig:diffusion}
\end{figure*}

In order to explain the experimental observations, we studied Co and Rh adatoms 
on fcc-Mn/Re(0001) using
density functional theory (DFT). 
First we discuss the structural and
magnetic properties of fcc-Mn/Re(0001).
From our DFT total energy calculations we find that the row-wise AFM state is by 300~meV/Mn atom lower than the ferromagnetic
(FM) state.
In the structural relaxations we have taken into account the symmetry breaking due to the row-wise AFM state highlighted by the experiments.
Surprisingly, we find
in addition to a relaxed Mn-Re interlayer distance, a lateral shift by $\sim$15~pm perpendicular to the 
$\uparrow\uparrow$-rows. Thereby, 
Re surface atoms are closer to the two adjacent nearest-neighbor Mn atoms with parallel magnetic moments ($m_{\rm Mn}\approx 3.4$~$\mu_{\rm B}$) with respect to the nearest-neighbor Mn atom with an opposite spin moment (Fig.\,\ref{fig:diffusion}a).
As a result the induced magnetic moments of the Re surface atoms
are enhanced, 
strengthening the antiferromagnetic Mn-Re interaction 
\footnote{The lateral shift of the Mn layer results in a total energy gain of 18 meV/Mn atom for the row-wise AFM state. Thereby, it becomes the magnetic ground state of the system since it is lower than all spin spiral states as well as the triple-Q state \cite{Spethmann2020}.}.
Note, that the magnetic moment direction of the Mn atoms is in-plane along the rows of parallel moments 
(Fig.\,\ref{fig:diffusion}a) due to the easy-plane magnetocrystalline anisotropy and the anisotropic symmetric exchange interaction \cite{Spethmann2020}.

To understand the experimentally found one-dimensional motion of Rh and Co atoms we calculate the minimum energy paths and the related energy barriers for different directions of atom movement using the climbing image nudged elastic band method (NEB)~\cite{neb1,neb2} (see Methods).
We begin with an adatom in its preferred adsorption site (I1), and move it to an equivalent site both along an $\uparrow\uparrow$-row and 
along an $\uparrow\downarrow$-row (Fig.\,\ref{fig:diffusion}a)
where the final states are indicated as F1 and F3, respectively,
and exhibit opposite spin directions.

We 
start our discussion with Rh atom movement 
along the $\uparrow\uparrow$-rows, i.e.\ 
from 
I1 to F3 (blue lines and blue transparent spheres in Figs.\,\ref{fig:diffusion}a,b). 
We find a symmetric minimum energy path with two identical saddle points
at the $\uparrow\downarrow$-bridge sites. The intermediate energy minimum along this path is the fcc hollow site which is by only $\approx 4$~meV unfavorable with respect to
the adsorption site I1
for a Rh atom. Note, that the induced magnetic moment of Rh deviates on the path only little from its value of $0.04$~$\mu_{\rm B}$ at I1.
The energy at the saddle points of $\sim$180 meV corresponds to the energy barrier for Rh atom movement along an $\uparrow\uparrow$-row. 

In contrast, the minimum energy path for Rh exhibits two saddle points of
different barrier height along an
$\uparrow\downarrow$-row, i.e.~from I1 to F1 (red lines and red transparent spheres in Fig.\,\ref{fig:diffusion}a,b).
The first saddle point results in a barrier of $\sim$240 meV and is 
due to the movement of the atom across the $\uparrow\uparrow$-bridge site. At this saddle point the induced Rh magnetic moment is much enhanced (0.3~$\mu_{\rm B}$). A second smaller energy barrier ($\sim$180 meV) occurs when the Rh atom starts from the intermediate state at the fcc hollow site and crosses the second saddle point, 
the $\uparrow\downarrow$-bridge site, to reach the final state F1. Therefore, the overall energy barrier to move a Rh atom 
along an $\uparrow\downarrow$-row amounts to $\sim$240 meV. 
These calculations are consistent with the favored movement of Rh atoms along $\uparrow\uparrow$-rows observed in our experiments. The higher barrier for Rh movement 
along $\uparrow\downarrow$-rows is related to a larger induced magnetic moment at the $\uparrow\uparrow$-bridge site and is thus of magnetic origin.

For the Co atom, we also find a
lower energy barrier for the path along the $\uparrow\uparrow$-rows,
i.e.~from I1 to F3 (blue transparent spheres and blue lines in Fig.\,\ref{fig:diffusion}a,c).
Due to exchange coupling, the Co magnetic moment is aligned parallel to the nearest-neighbor Mn atoms both at position I1 and F3. Along the minimum energy path the Co moment remains along this direction, however, its magnitude varies slightly. The Co atom crosses two identical saddle points given by $\uparrow\downarrow$-bridge sites resulting in an energy barrier height of $\sim$150 meV.

For the minimum energy path of a Co atom along an $\uparrow\downarrow$-row (red transparent spheres and red lines in Fig.\,\ref{fig:diffusion}a,c), its magnetic moment must flip from $+1.6$~$\mu_{\rm B}$ at I1 to $-1.6$~$\mu_{\rm B}$ at F1 in order to be in the energetically favorable spin configuration in both initial and final state. As for Rh, we observe two different energy barriers along this path with a lower barrier of $\sim$110 meV at the $\uparrow\uparrow$-bridge site and a more than twice as high barrier of $\sim$260 meV at the $\uparrow\downarrow$-bridge site. Note, that at the latter saddle point the Co magnetic moment is reduced to $0.4$~$\mu_{\rm B}$
which results from the required spin-flip along the path and the constraint to collinear magnetic moments in our calculation.
With an overall barrier height of $\sim$260 meV, this path is clearly unfavorable with respect to Co movement along $\uparrow\uparrow$-rows consistent with our experimental observations.


To reveal the role of the Co spin-flip on the energy barrier for the path along $\uparrow\downarrow$-rows, we have performed a minimum energy path calculation from I2 to F2 ({\tq} transparent spheres and line in Fig.\,\ref{fig:diffusion}a,c). Along this path the magnetic moment of the Co atom stays in the same direction
and a large reduction of the
magnetic Co moment on the $\uparrow\downarrow$-bridge site is avoided. 
Since the initial state I2 of this path coincides with the intermediate energy minimum along the path from I1 to F1 (red line in Fig.\,\ref{fig:diffusion}c), we can combine the first half of the path I1-F1 with the path I2-F2 to obtain a minimum energy path from I1 to F2. On this path along an $\uparrow\downarrow$-row the magnetic moment of the Co atom does not flip
and it amounts to 1.5~$\mu_{\rm B}$ at the $\uparrow\downarrow$-bridge site. The
spin of the Co atom may flip after the metastable state F2 has been
reached such that a relaxation into the favorable state F1 is
achieved. Nevertheless,
the energy along the path from I1 to F2 without a spin-flip
is asymmetric with two saddle points of different energy as for the I1-F1 path with a spin flip 
(red line in Fig.\,\ref{fig:diffusion}c). Since the energy barrier due to the saddle point between I2 and F2 is higher than that for the path along the $\uparrow\uparrow$-rows ({\tq} vs.~blue line in Fig.\,\ref{fig:diffusion}c) the latter remains the favorable path for Co atom motion.

\section{Discussion}

The first-principles calculations nicely explain the experimentally observed moving direction of Co and Rh atoms by anisotropic potential landscapes arising from the row-wise AFM state.
The spontaneous magnetic ordering causes an electronic anisotropy, which drives the $\sim$15~pm lateral shift of the Mn layer relative to the Re(0001) surface. The resulting similarity to diffusion on surfaces with low structural symmetry such as Ag(110) \cite{Li1998} or Cu(211) \cite{Meyer1995} is striking. However, there are two important differences: antiferromagnetic domains can be switched by lateral currents \cite{Baltz2018}, which in principle allows to externally control diffusion directions on antiferromagnetic surfaces. Furthermore, since the Mn layer itself remains structurally hexagonal, the $\uparrow\uparrow$-rows are not necessarily the easy diffusion direction for all atoms or molecules. We have performed 
additional DFT calculations which show for instance that Cu atoms behave inversely to Rh atoms, i.e.\ the energy barrier at the $\uparrow\uparrow$-bridge is lower than at the $\uparrow\downarrow$-bridge, which hinders Cu diffusion along the $\uparrow\uparrow$-rows.

The experimentally found higher mobility, i.e.\ longer travel distances
of Co atoms, agrees qualitatively with a lower barrier of $\sim$150~meV compared to $\sim$180~meV for Rh.
Additional effects beyond this adiabatic analysis might contribute to the high mobility of Co. Firstly, our DFT calculations show a total binding energy of Co atoms on Mn/Re(0001) which is 1.5~eV lower than for Rh atoms. This means that a ballistic movement at an increased distance from the surface \cite{Guo2015} requires less energy for Co atoms than for Rh atoms. Since in the experiment some of the energy of the tunneling electrons is used for the excitation, the excess energy could be larger for the Co atoms. This is in accordance with the larger threshold voltage observed for Rh atoms.
Secondly, the $3d$ orbitals of Co have a smaller extent than the $4d$ orbitals of Rh, which may result in a reduced interaction with the surface and a reduced damping for a moving Co atom. Thirdly, Rh is about 75\% heavier. The lighter and faster moving Co atom can therefore cover more distance before it comes to rest. This argument only holds for damping processes which scale with travel time \cite{Brune1992}. Finally, the spin degree of freedom might contribute to the travel distance. Long jumps have been found experimentally for molecules \cite{Hla2004, Civita2020} and one reason is their ability to store energy during movement in form of vibrations. Spin excitations of the Co atom and within the Mn layer might play a similar role here.

In conclusion, the observed one-dimensional movement demonstrates that magnetism can play a decisive role in controlling atom movement, with possible consequences for related phenomena such as nanostructure growth, molecular self-assembly and catalysis.


\section{Methods}
\subsection{Experimental details}   
The experiments were performed in a multi-chamber ultra-high vacuum system with different chambers for substrate cleaning, metal deposition, and STM measurements. The Re(0001) single crystal surface was cleaned by cycles of annealing in an oxygen atmosphere of $10^{-7}$--$10^{-8}$~mbar at temperatures of up to $T=1400$~K; before metal deposition a final flash to $T = 1800$~K was performed. The Mn was evaporated from a pyrolytic boron nitride (PBN) Knudsen cell of volume 1 cm$^3$ and held at $T = 670$~$^\circ$C, resulting in a flux of approximately 0.1 atomic layers per minute; during Mn deposition the Re single crystal was still at elevated temperature ($\approx 100$~$^\circ$C) from the final flash.

The Mn/Re(0001) samples were then cooled to $T=4.2$~K within the STM. After a quick transfer to the e-beam evaporator ($\Delta t< 5$~s) Co or Rh atoms were evaporated from a rod of 2 and 1 mm diameter, respectively, at a rate of about 0.1 atomic layers per minute. We estimate a surface temperature of $T=25$\,K~$\pm 5$\,K during adatom deposition.

We use a Cr bulk STM tip, etched in 1M HCl solution. Within the STM the tip was cleaned by field emission on a W(110) surface and sharpened by voltage pulses of $U=4$--6~V and gentle surface collisions on the Mn/Re(0001) sample. Except for the measurement shown in Fig.\,\ref{fig:electronic}a,b we did not optimize the tip for magnetic contrast. Consequently, in all other data the tip exhibits no significant spin sensitivity at the used tunnel voltages, thereby simplifying the data analysis.
Magnetic domain walls were imaged by making use of their (spin-averaged) electronic signal \cite{Spethmann2021} and the orientation of rotational domains was determined via the symmetry of native defects: Extended Data Fig.\,\ref{fig:methods} shows a surface area with all three rotational domains, demonstrating the strict correlation of domain orientation and defect shape orientation.

For atom ``kicking'', the Co or Rh atom was first moved to a defect-free area and the tip was then stabilized precisely above the atom, using an atom-tracking feedback by steepest ascent \cite{nanonis} and tunnel parameters on the order of $U=+10$~mV and $I=200$~pA. The atom-tracking and height-control feedback were then switched off and the voltage raised abruptly at constant tip height for $\Delta t=0.4$--1.0~s to $U_\text{P}=\pm 200$~mV and $U_\text{P}=+1$~V, respectively. A voltage pulse is thus automatically accompanied by an increased current through the adatom, but only as long as the atom does not move. For the measurement series displayed in Fig.\,\ref{fig:Co}f, the tip was stabilized above the atoms at $U=+10$~mV and $I=200$~pA, and after the kick the atoms were moved back to their original positions using constant current pulling mode at approximately $U=4$~mV and $I=60$~nA. Since micro-tip changes are more likely during manipulation than during imaging, the necessity to move an atom back into position, due to surrounding defects, limits the number of kicks which can be performed with the same micro-tip. Which of the two equivalent directions along the $\uparrow\uparrow$-rows is taken, depends on the specific micro-tip: for some tips, like the one used in Fig.\,\ref{fig:Co}a--e, one movement direction is favored; other tips show a more symmetric distribution. The data of Fig.\,\ref{fig:Co} has been smoothed by a 3-pixel Gauss filter; all other figures show raw data.

\subsection{Computational details}
\label{sec:comp_details}
Our first-principles calculations are based on density-functional theory (DFT) within the projector augmented wave (PAW)~\cite{Blochl1994} 
method as implemented in the \textsc{vasp} code~\cite{Kresse1996,Kresse1999}.
We have performed spin-polarized calculations using Wigner-Seitz radii for the elements,
i.e. $R_{\rm Mn}^{\rm WS} = 1.32$~\AA\, for 
Mn and $R_{\rm Re}^{\rm WS} = 1.43$~\AA\, 
for Re. As lattice parameters we used $a_{NN} = 2.78$~\AA\,, 
$c =4.49$~\AA, 
which were taken from Ref.~\onlinecite{Ji2016} obtained via DFT within the generalized gradient approximation (GGA).
Films with fcc-stacking of the Mn monolayer were structurally relaxed using the GGA exchange-correlation (xc) potential~\cite{Perdew1996} in the
row-wise AFM state using an asymmetric film consisting of six Re(0001) layers with a Mn layer on top. The Mn layer and the uppermost Re layer were relaxed in $x$, $y$, and $z$-direction, where the bulk reference is $c/2 = 2.24$~\AA. 
We used $5 \times 5 \times 1$ $k$~points in the irreducible wedge of the two-dimensional (2D) Brillouin zone (BZ) and a plane wave basis
set cutoff of $E_{max} = 350$~eV. The structures were optimized using the conjugate gradient method with forces calculated from the Hellman-Feynman theorem.
Structures were considered to have been optimized when all the forces were smaller than 0.01 eV/{\AA}. We have used a $(5\times 6)$ super cell in order to design the row-wise AFM state. 

The Co atom preferentially adsorbs in the hcp hollow 
site (indicated by I1), which is about 50~meV lower in energy compared to the fcc hollow site (I2). In both cases a small lateral shift relative to the center of the Mn atom triangle is found. The configuration of the three nearest Mn magnetic moments results in a net ferromagnetic coupling 
to the Co magnetic moment ($1.6$~$\mu_{\rm B}$),
where two of them are parallel and one is antiparallel to the Co moment (Fig.\,\ref{fig:diffusion}a). 
For the Rh atom the two hollow sites are almost energetically degenerate, with a difference of only $\sim$4 meV, and in the hcp site, the adjacent Mn atoms induce a Rh magnetic moment of $\sim$0.04 $\mu_{\rm B}$.

The magnetic moment of the Co adatom at hcp hollow site F1 (Fig.\,\ref{fig:diffusion}a) is $-1.6$ $\mu_{\rm B}$ due the magnetic moment direction of the two nearest-neighbor Mn moments (net ferromagnetic coupling). 
In order to obtain a net antiferromagnetic coupling to the Co magnetic moment with the three nearest neighbor Mn atoms, 
we perturbed the position of the Co adatom by a small amount ($\sim$0.05 {\AA}) and relaxed the structure with an initial high value of 
positive magnetic moment for the Co adatom (net antiferromagnetic coupling with the three nearest neighbor Mn atoms). 
Upon structural relaxation, we find that the Co atom moved to position F2 which is shifted by 15~pm with respect to F1 in the direction perpendicular to the $\uparrow\uparrow$-rows. At this site the magnetic moment of the Co atom is $+1.2$ $\mu_{\rm B}$ with a net antiferromagnetic coupling to the three nearest neighbor Mn atoms.  
Our calculation shows that F2 is a higher energy local minima state by $\sim$40 meV compared to the F1 state with an opposite spin moment.

We calculate the energy barriers and the minimum
energy paths between two known adatom adsorption sites by using the climbing image nudged elastic band method (NEB)~\cite{neb1,neb2}. In the NEB method we optimize a number of intermediate images to their lowest energy possible while maintaining equal spacing to neighboring images. This is obtained by adding a spring force along the band between the images. The climbing image method is a small modification to the NEB method where the highest energy image is pushed up to the saddle point to get the exact position of the saddle point. This is done by inverting the true force for this image along the tangent. Thus this image does not feel the spring force. In our NEB calculations we have not relaxed the surface atoms for simplicity. We have used a mixture of quick-min and LBFGS  optimizers~\cite{neb_optimizer} and $-$5.0 eV/{\AA$^2$} spring force for our calculations.     



For the simulated STM and SP-STM images of Mn/Re(0001),
we have used the full-potential linearized augmented plane wave method (FLAPW) as implemented in the {\tt FLEUR} code~\cite{FLEUR}. We have performed spin-polarized calculations using radii for the muffin tin spheres (in all {\tt FLEUR} calculations), i.e. $R^{\rm Mn}_{\rm MT}$ = 2.35 a.u. for Mn and $R^{\rm Re}_{\rm MT}$ = 2.40 a.u. for Re. 
For the {\tt FLEUR} calculations of Mn/Re(0001) we have used asymmetric films with 6 layers of Re substrate. Here we used 1444 $k$ points in the full Brillouin zone and an energy cutoff of $4.3\,\rm{a.u.}^{-1}$.

\subsection{Simulated STM/SP-STM images of Mn/Re(0001)}
\label{sec:stm_simulation}
A comparison of simulated STM and SP-STM images with and without a shift of the pseudomorphic Mn layer in the row-wise AFM state is shown in Extended Data Fig.\,\ref{fig:stm_below} and \ref{fig:stm_above} for a small energy window below and above the Fermi level, respectively. The STM and SP-STM images are calculated based on the electronic structure from DFT 
using the Tersoff-Hamann model \cite{Tersoff1983}
and its extension to spin-polarized STM \cite{Wortmann2001}.

The simulated SP-STM images
(Extended Data Fig.\,\ref{fig:stm_below}b,d and \ref{fig:stm_above}b,d) show the expected stripe period of the row-wise AFM state observed experimentally in Fig.\,\ref{fig:electronic}a. This magnetic contrast
is the same for the unshifted and for the
laterally shifted Mn layer.
In the simulated STM images for a positive bias voltage (Extended Data Fig.\,\ref{fig:stm_above}c), we find for the laterally shifted Mn 
layer also the experimentally observed stripes of atomic period shown in Fig.\,\ref{fig:electronic}d. This stripe contrast is of electronic origin and it is
absent in the simulated STM images of the
unshifted Mn layer for both positive and negative bias voltages (Extended Data Fig.\,\ref{fig:stm_below}a and \ref{fig:stm_above}a).
This supports that the lateral shift of the Mn layer predicted by our DFT total energy calculations occurs in the experiment.

Note, that in the experimental SP-STM image shown in Fig.\,\ref{fig:electronic}b, the electronic and magnetic stripes are not in phase as in Extended Data Fig.\,\ref{fig:stm_above}c,d. However, in the experiment of Fig.\,\ref{fig:electronic}b the tunneling
current was extremely large, i.e.~the tip-sample distance was very small, which might lead to additional effects due to tip-sample
interaction. Another possible reason for the difference
with this particular image may be specific micro-tip properties.


\subsection{Co electric dipole and voltage polarity}
\label{sec:efield}
Concerning voltage polarity, the data of Fig.\,\ref{fig:Co}f shows a trend to larger kicking distances for negative polarity, a much weaker effect than observed for other systems \cite{Whitman1991,Stroscio1991,Civita2020}. In further experiments with $U_\text{P}=\pm 500$~mV we found no statistically significant effect of polarity on travel distance, indicating that the local electric field does not dominate the Co atom movement. In addition, other effects can cause a voltage polarity dependence, for instance an asymmetric density of states with respect to the Fermi level. Indeed, the vacuum density of states of Mn/Re(0001) is about four times higher 200~meV below compared to 200~meV above the Fermi level (see Fig.\,S9 in Ref.\,\onlinecite{Spethmann2020}). This leads to a higher current flowing at pulses of $U_\text{P}=-200$~mV compared to $U_\text{P}=+200$~mV, which might explain the polarity trend observed in Fig.\,\ref{fig:Co}f.

To further investigate the impact of the tip's local electric field onto the atom movement we have calculated the Co electric dipole using the Bader charge analysis \cite{Bader}. It turns out that the value is sizable with $\mu=0.61$~e\AA\ or 2.9 D; for comparison, H$_2$O has an electric dipole of $\mu=1.85$~D. When a charge-neutral Co atom is placed on the Mn layer, electrons are transferred from the surface toward the Co atom, producing an electric dipole pointing down. Neglecting work function differences between tip and sample, at positive (negative) voltage the tip's electric field points up (down), decaying with lateral distance from the tip position. Thus, a repulse (attractive) force should act on the Co atom at positive (negative) voltage. This means that the weak polarity trend observed in Fig.\,\ref{fig:Co}f cannot be explained by the tip's electric field. Thus, despite the sizable Co dipole, electric fields do not play a role in the experiments.

\section{Data availability}
The STM and DFT data are available upon reasonable request.

\section{Code availability}
The STM data was analyzed with the open access software Gwyddion (http://gwyddion.net). 

\section{Acknowledgements}
We acknowledge funding from the Deutsche Forschungsgemeinschaft (DFG, German Research Foundation) under project numbers 408119516, 
418425860, and 445697818. 
This work was performed using HPC resources from the North-German Supercomputing Alliance (HLRN).

\section{Author contributions}
A.K.\ devised the experiments. A.K.\ and F.Z.\ prepared the samples, performed the STM experiments, and together with K.v.B.\ analyzed the experimental data. S.H.\ performed and analyzed the calculations with St.H. The figures were prepared by A.K., F.Z., and S.H. The manuscript was written by A.K., S.H., and St.H., with all authors contributing.

\newpage

\setcounter{figure}{0}

\begin{figure}
    \centering
    \renewcommand{\figurename}{Extended Data Fig.}
    \includegraphics[width=0.5\linewidth]{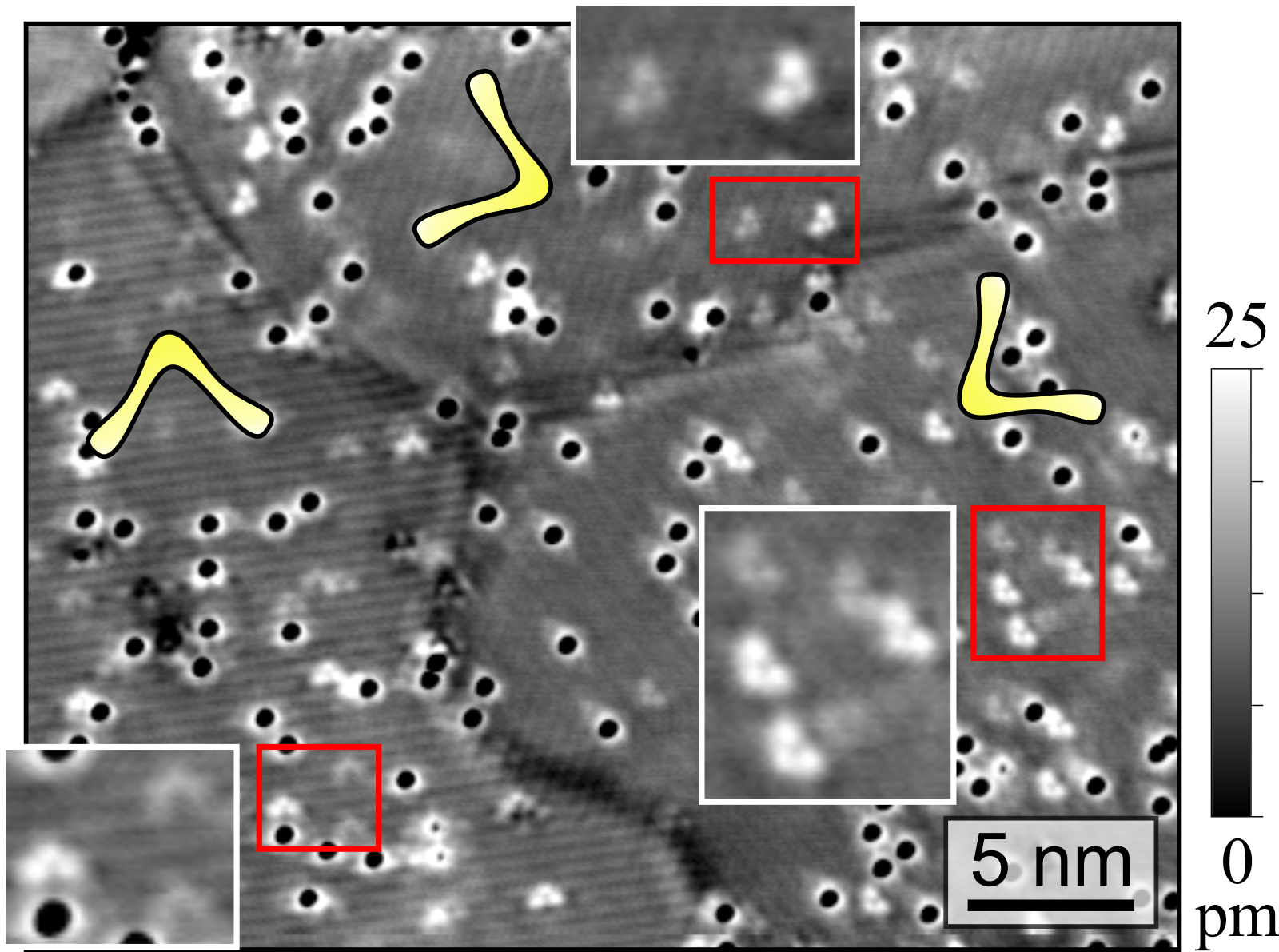}
    \caption{\textbf{| Correlation of row-wise AFM state and defect shape.} Intersection of three rotational domains, two of them showing an almost vanishing spin contrast ($U = -30$~mV, $I = 7$~nA). Some defect types display a broken $C_3$ symmetry, correlated with the direction of the $\uparrow\uparrow$-rows, compare zoom-ins and yellow boomerangs. This correlation allows to determine the orientation of row-wise AFM domains in non-magnetic STM measurements.}
    \label{fig:methods}
\end{figure}

\begin{figure}[htbp]
	\centering
	\renewcommand{\figurename}{Extended Data Fig.}
    \includegraphics[width=0.5\linewidth]{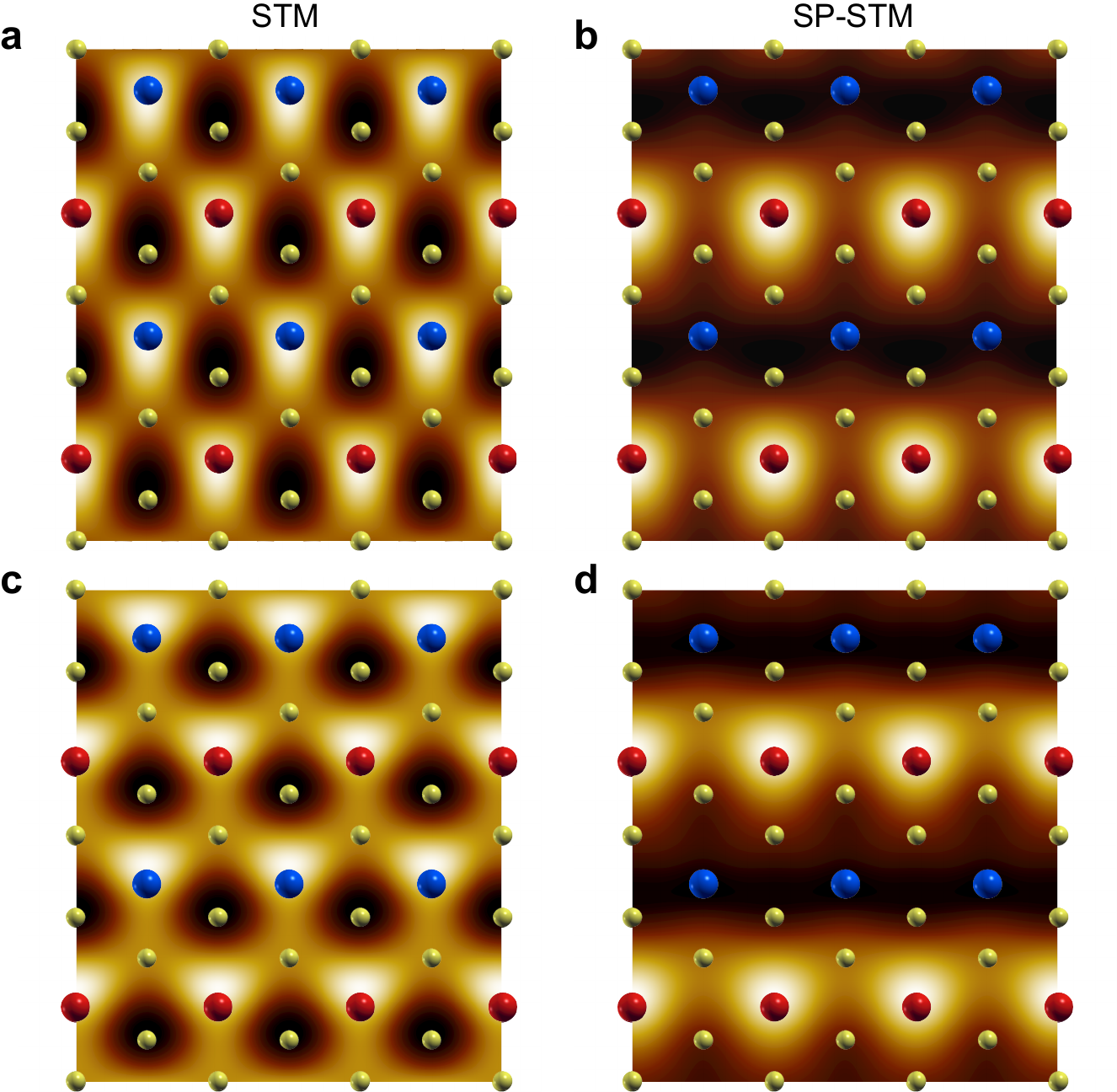}
    \caption{\textbf{| STM simulations for Mn/Re(0001) at a bias voltage of $U=-$50~mV.} Comparison of simulated STM and SP-STM images of fcc-Mn/Re(0001)
    in the row-wise AFM state based on DFT calculations 
    at a bias voltage of $-$50 meV and a distance of 3 {\AA} above the surface without and with a lateral shift of the Mn monolayer. \textbf{a,b} Mn atoms are at perfect fcc positions. \textbf{c,d} Mn atom rows shifted from the perfect fcc sites into the energetically favorable
    relaxed positions.  Red and blue spheres represent Mn atoms of opposite spin directions, while yellow spheres denote Re atoms.
    Left panels show STM images and right panels show SP-STM images assuming a tip spin-polarization of 0.5.} 
    \label{fig:stm_below}
\end{figure}

\begin{figure}[htbp]
	\centering
	\renewcommand{\figurename}{Extended Data Fig.}
    \includegraphics[width=0.5\linewidth]{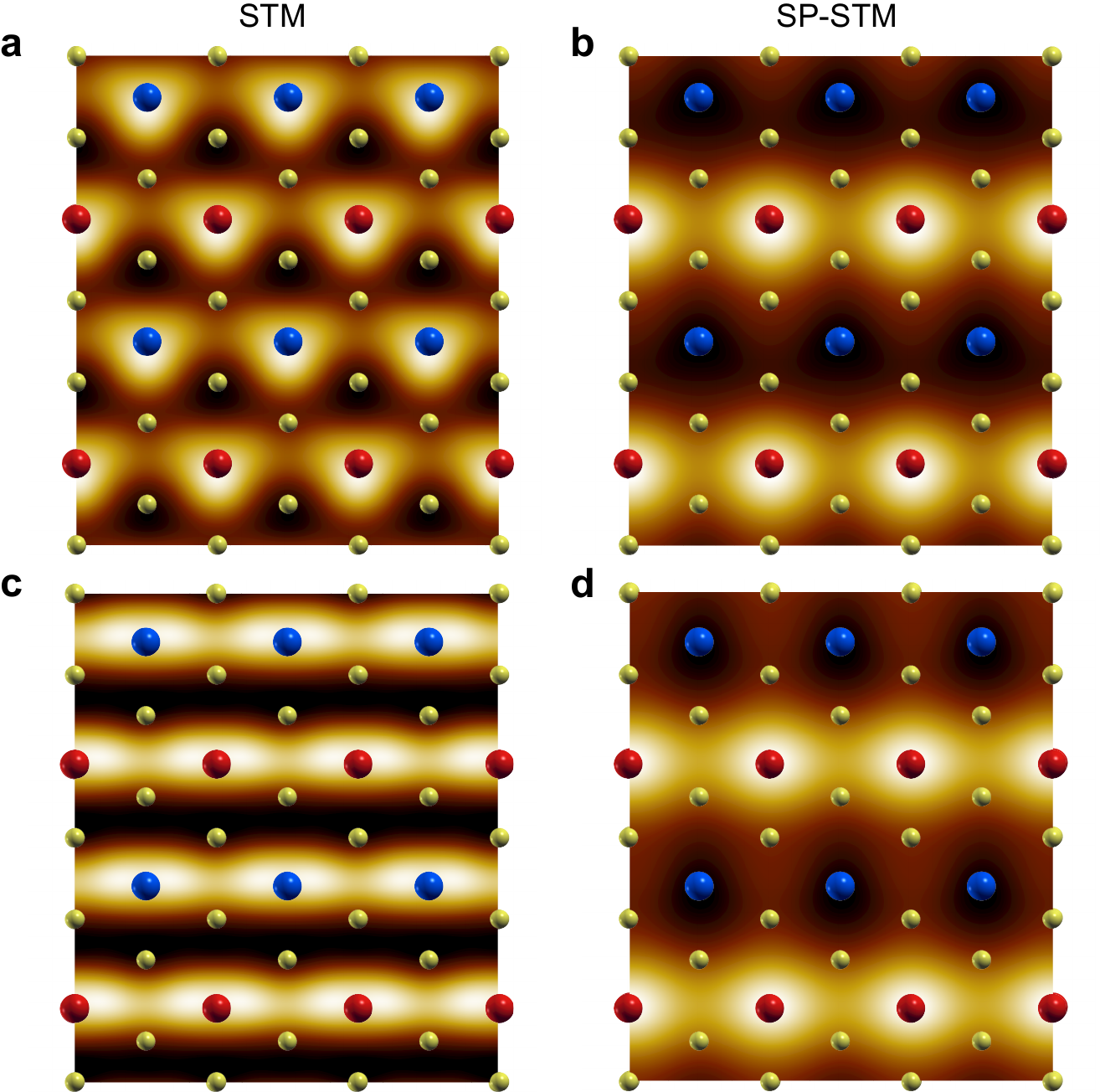}
    \caption{\textbf{| STM simulations for Mn/Re(0001) at a bias voltage of $U=+$50~mV.} Comparison of simulated STM and SP-STM images of fcc-Mn/Re(0001)
    in the row-wise AFM state based on DFT calculations 
    at a bias voltage of $+$50 meV 
    and a distance of 3 {\AA} above the surface
    without and with a lateral shift of the Mn monolayer. \textbf{a,b} Mn atoms are at perfect fcc positions. \textbf{c,d} Mn atom rows shifted from the perfect fcc sites into the energetically favorable
    relaxed positions.  Red and blue 
    spheres represent Mn atoms of opposite spin 
    directions, while yellow spheres denote Re atoms.
    Left panels show STM images and right panels show SP-STM images assuming a tip spin-polarization of 0.5.
    } 
    \label{fig:stm_above}
\end{figure} 

\end{document}